\documentclass[table,aps,prl,reprint,superscriptaddress, showpacs]{revtex4-2}

\usepackage{amssymb}
\usepackage{amsmath}
\usepackage{graphicx}
\usepackage{natbib}
\usepackage{xcolor}

\usepackage{hyperref}

\usepackage{floatrow}
\newfloatcommand{capbtabbox}{table}[][\FBwidth]
\usepackage{blindtext}

\DeclareUnicodeCharacter{2212}{-}
\DeclareUnicodeCharacter{00A0}{ }
\DeclareUnicodeCharacter{0308}{HERE!HERE!}

\newcommand{\RN}[1]{%
  \textup{\uppercase\expandafter{\romannumeral#1}}%
}

\begin{document}

\preprint{APS/123-QED}

\title{Coherent control of a few-channel hole type gatemon qubit}

\author{Han Zheng}
\email{han.zheng@unibas.ch}
\affiliation{Quantum- and Nanoelectronics Lab, Department of Physics, University of Basel, 4056 Basel, Switzerland}
\author{Luk Yi Cheung}
\affiliation{Quantum- and Nanoelectronics Lab, Department of Physics, University of Basel, 4056 Basel, Switzerland}
\author{Nikunj Sangwan}
\affiliation{Quantum- and Nanoelectronics Lab, Department of Physics, University of Basel, 4056 Basel, Switzerland}
\author{Artem Kononov}
\affiliation{Quantum- and Nanoelectronics Lab, Department of Physics, University of Basel, 4056 Basel, Switzerland}
\author{Roy Haller}
\affiliation{Quantum- and Nanoelectronics Lab, Department of Physics, University of Basel, 4056 Basel, Switzerland}
\author{Joost Ridderbos}
\affiliation{MESA+ Institute for Nanotechnology University of Twente, 7500 AE Enschede, The Netherlands}
\author{Carlo Ciaccia}
\affiliation{Quantum- and Nanoelectronics Lab, Department of Physics, University of Basel, 4056 Basel, Switzerland}
\author{Jann Hinnerk Ungerer}
\affiliation{Quantum- and Nanoelectronics Lab, Department of Physics, University of Basel, 4056 Basel, Switzerland}
\author{Ang Li}
\affiliation{Department of Applied Physics, Eindhoven University of Technology, 5600 MB Eindhoven, The Netherlands}
\author{Erik P.A.M. Bakkers}
\affiliation{Department of Applied Physics, Eindhoven University of Technology, 5600 MB Eindhoven, The Netherlands}
\author{Andreas Baumgartner}
\email{andreas.baumgartner@unibas.ch}
\affiliation{Quantum- and Nanoelectronics Lab, Department of Physics, University of Basel, 4056 Basel, Switzerland}
\affiliation{Swiss Nanoscience Institute, University of Basel, 4056 Basel, Switzerland}
\author{Christian Sch{\"o}nenberger}
\email{christian.schoenenberger@unibas.ch}
\affiliation{Quantum- and Nanoelectronics Lab, Department of Physics, University of Basel, 4056 Basel, Switzerland}
\affiliation{Swiss Nanoscience Institute, University of Basel, 4056 Basel, Switzerland}

\date{\today}

\begin{abstract}

Gatemon qubits are the electrically tunable cousins of superconducting transmon qubits. In this work, we demonstrate the full coherent control of a gatemon qubit based on hole carriers in a Ge/Si core/shell nanowire, with the longest coherence times in group IV material gatemons to date. The key to these results is a high-quality Josephson junction obtained in a straightforward and reproducible annealing technique. We demonstrate that the transport through the narrow junctions is dominated by only two quantum channels, with transparencies up to unity. This novel qubit platform holds great promise for quantum information applications, not only because it incorporates technologically relevant materials, but also because it provides new opportunities, like an ultra-strong spin-orbit coupling in the few-channel regime of Josephson junctions.

\end{abstract}

\maketitle

\section{Introduction}

Quantum computing has become a topic of intense activity due to its potential to revolutionize information processing \cite{brooks2023quantum}. Presently, one of the most popular platforms are superconducting circuits in the form of transmon qubits, which are already employed in solving noisy intermediate-scale quantum problems \cite{preskill2018quantum,arute2019quantum,kjaergaard2020superconducting}. Conventional transmons rely on metallic superconductor-insulator-superconductor (SIS) tunnel junctions, in which tuning the qubit frequency  requires a magnetic flux in a SQUID geometry \cite{krantz2019quantum}. However, flux cross-talk between qubits and the heat load due to the flux control currents are problematic for scaling up the number of qubits \cite{krinner2019engineering}. 

An alternative to metallic transmons are semiconductor-superconductor (Sm-S) hybrid systems \cite{de2010hybrid}, so-called gatemon qubits \cite{larsen2015semiconductor,de2015realization}, with a wealth of recently demonstrated related concepts, like parity protected qubits \cite{larsen2020parity}, gate-tunable fluxonium qubits \cite{pita2020gate}, gatemons operated with only a few highly transparent quantum channels \cite{kringhoj2018anharmonicity,danilenko2023few}, or Andreev level \cite{zazunov2003andreev,janvier2015coherent,cheung2023photon} and Andreev spin qubits \cite{tosi2019spin,hays2020continuous,hays2021coherent,pita2023direct}. All these devices rely on high-quality crystals with near-perfect Sm-S interfaces. This has limited the material choice mainly to InAs-based systems, like vertically-grown InAs NWs \cite{larsen2015semiconductor,de2015realization,casparis2016gatemon,luthi2018evolution,danilenko2023few}, InAs 2D systems \cite{casparis2018superconducting}, 
or selective-area-grown InAs NWs \cite{hertel2022gate}. However, III/V materials are difficult to integrate in standard CMOS technologies, and hyperfine interactions introduce additional decoherence.

More suitable for CMOS technology would be group IV materials, but only very few gatemon-related experiments have been reported so far: on carbon-nanotubes \cite{mergenthaler2021circuit}, graphene \cite{wang2019coherent}, and large-diameter Ge/Si core/shell NWs \cite{zhuo2023hole}, with the corresponding qubit coherence times more than an order of magnitude shorter than in III/V materials. However, improvements are expected, especially for the technologically relevant CMOS compatible Si and Ge systems \cite{gunn2005methods,veldhorst2017silicon}. In particular, Ge is of increasing interest for quantum information processing \cite{scappucci2021germanium}, because hole states in Ge have a p-wave symmetry, inherently reducing hyperfine interactions, with further improvements expected from isotopic purification. Even more promising might be Ge/Si core/shell nanowires with a 1D hole gas strongly confined by the Ge/Si interface \cite{xiang2006ge,conesa2017boosting} and an electrically tunable, very large “direct” Rashba spin-orbit interaction \cite{kloeffel2018direct,froning2021ultrafast}. 

In this work, we report the full functionality of a gatemon qubit based on narrow Ge/Si core/shell NWs. The key step is the fabrication of highly transparent Josephson junctions (JJs), in an undemanding ex-situ annealing step \cite{weber2017silicon,heinzig2012reconfigurable} driving a thermally activated propagation of superconducting aluminum into the Ge NW core  \cite{ridderbos2018josephson,sistani2019highly,delaforce2021ge}. We incorporate such JJs in a gatemon qubit device and explicitly demonstrate the electrical tunability of the qubit frequency and the full coherent  control of the qubit in the time domain, with an energy relaxation time on par with III/V systems. Most importantly, these experiments allow us to analyse the qubit anharmonicity, suggesting that the supercurrent through the junction is dominated by two ballistic, virtually Schottky-barrier free conductance channels. This work establishes narrow Ge/Si core/shell NWs as a promising Sm-S hybrid platform for quantum information processing, and opens new avenues to investigate novel effects, for example, in circuit-quantum-electrodynamics (circuit QED) experiments on NWs \cite{nadj2010spin,petersson2012circuit} with ultra-strong spin-orbit interactions.

\section{Device}

\begin{figure}[b]
  \centering
  \includegraphics[width=\linewidth]{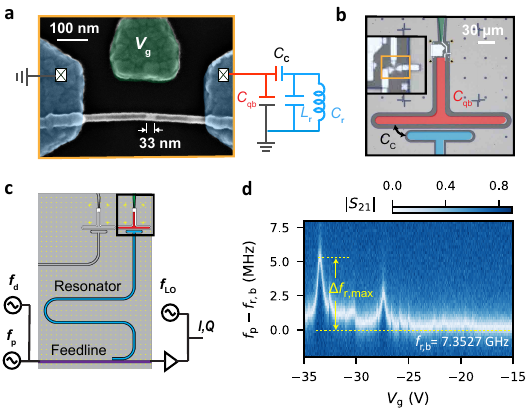}
  \caption{\textbf{Ge/Si core/shell nanowire based gatemon device.} \textbf{a}, False-colored SEM image of the Al-Ge-Al junction. The bright segment of the NW has a Ge core, while the dark segments contains Al. The side gate for electrical qubit control is colored in green. The readout resonator is shown as a simplified equivalent circuit on the right. \textbf{b}, False-colored optical micrograph of a nanowire Josephson junction shunted by a T-shaped capacitor (red) to the surrounding ground plane and capacitively coupled ($C_c$)  to a superconducting $\lambda/4$ resonator (blue). \textbf{c}, Overview of the circuit QED chip and schematic of the readout and control circuit. The $\lambda/4$ resonator is inductively coupled to a feedline for readout. \textbf{d}, Transmission amplitude $\lvert S_{21} \rvert$ through the feedline as a function of the gate voltage $V_{\rm g}$ and the probe frequency $f_{\rm p}$, plotted as the difference from the bare resonator frequency $f_{\rm r,b}$. A bright signal signifies the resonator frequency $f_{\rm r}$.}
  \label{fig:1}
\end{figure}

Our gatemon qubit can be understood as a non-linear LC oscillator that consists of a gate-tunable Ge/Si core/shell NW JJ as the non-linear inductance, depicted in Fig.~\ref{fig:1}a, and a shunt capacitor $C_{\rm qb}$ shown in Fig.~\ref{fig:1}b. The NW has a diameter of $\sim 20\,$nm, and is expected to have minimal strain-induced defects and a high carrier mobility due to the [110] growth direction \cite{conesa2017boosting}. Such a NW was transferred to an undoped Si/SiO$_2$ substrate using a micro-manipulator, and contacted by Al using standard lift-off techniques. The crucial step in forming the highly transparent JJ is a thermal annealing step at $200\,^{\circ}$C, which drives an inter-diffusion process of Al atoms from the contacts into the Ge core, yielding an Al-Ge-Al junction with an atomically sharp interface  \cite{sistani2019highly}. The Si shell remains intact during this process. The bright NW segment in the electron micrograph of Fig.~\ref{fig:1}a has a Ge core, while the Al-filled segments show a darker contrast. The length of the Ge segment can be controlled by the annealing time, and directly read off as $33\,$nm for this specific device. The interface between Al and Ge turns out to be highly transparent, since it allows for coherent transport of Cooper pairs through the semiconducting Ge, resulting in a JJ \cite{ridderbos2019hard,ridderbos2018josephson}. The Josephson energy $E_{\rm J}$ can be tuned by a side gate voltage $V_{\rm g}$.

The gatemon shunt capacitance $C_{\rm qb}$ (red) is provided by a T-shaped NbTiN island etched into the surrounding ground plane, as shown in Fig.~\ref{fig:1}b. From electromagnetic simulations, we estimate $C_{\rm qb}\approx 78\,$fF. The island is galvanically connected to the JJ, forming the non-linear LC-oscillator, with the lowest two oscillator states forming the qubit states $\lvert 0 \rangle$ and $\lvert 1 \rangle$. The corresponding qubit frequency is determined by the gate tunable Josephson energy $E_{\rm J}$ and the charging energy $E_{\rm c}=e^2/2C_{\rm qb}\approx h\cdot 248\,$MHz. The charging energy is fixed and designed to be much smaller than the typical Josephson energy, which significantly reduces the sensitivity to charge noise \cite{koch2007charge}. In addition, we deposited a $100\,$nm thick Al layer on the surrounding NbTiN ground plane as a quasiparticle trap (see SI Fig.~\ref{figsup:Al traps}). 

An overview of the circuit QED chip is shown in Fig.~\ref{fig:1}c. The qubit state is probed via a capacitively coupled $\lambda /4$ coplanar transmission line resonator. The short end of the microwave resonator with a current anti-node is inductively coupled to the feedline, while the open end with a voltage anti-node is capacitively coupled to the qubit island, forming the qubit-resonator coupling capacitance $C_{\rm c}$. This device was measured in a dilution refrigerator with a base temperature of $15\,$mK, with consistent results in three different cooldowns.

\section{Measurements}

\subsection{Gate tunable qubit frequency with a two-channel Josephson junction}

\begin{figure*}[ht]
  \centering
  \includegraphics[width=\linewidth]{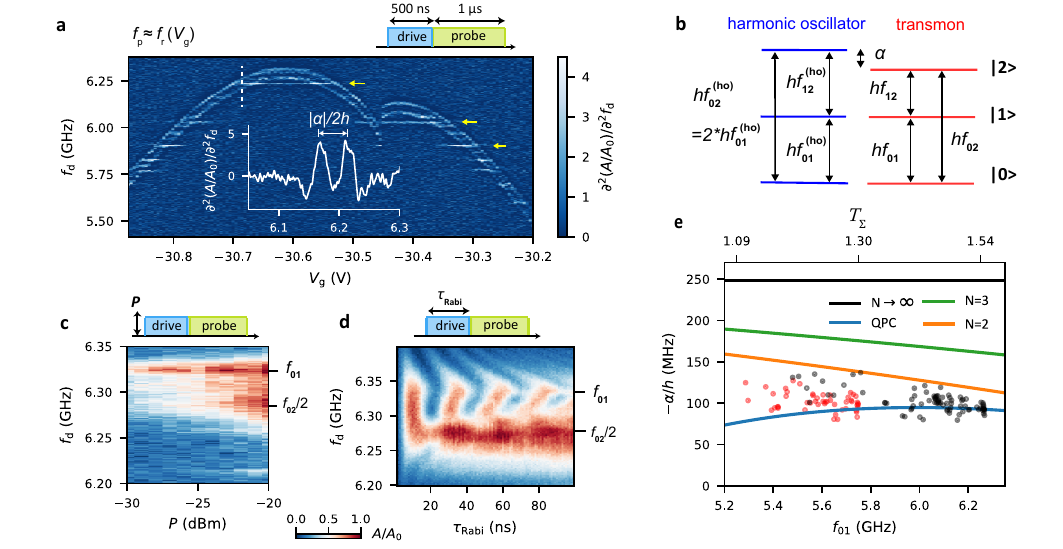}
  \caption{ \textbf{Qubit spectroscopy and anharmonicity.} \textbf{a}, Second derivative of the normalized resonator response $A/A_0$ with respect to $f_{\rm d}$ in a two-tone spectroscopy experiment as a function of the $f_{\rm d}$ and $V_{\rm g}$. The pulse sequence is shown schematically on the top right. The drive power $P$ is set to $-20\,$dBm and the probe frequency close to the resonator frequency $f_{\rm r}$. The inset shows a cross section at $V_{\rm g}=-30.7\,$V, clearly showing two peaks. \textbf{b}, Energy diagram of a transmon qubit showing explicitly the anharmonicity $\alpha$. \textbf{c,d}, Normalized resonator response $A/A_{0}$ in a two-tone spectroscopy experiment as a function of the drive frequency $f_{\rm d}$ and the drive power $P$, and as a function of $f_{\rm d}$ and the drive pulse duration $\tau_{\rm Rabi}$, respectively. The latter exhibits characteristic Rabi oscillations. \textbf{e}, Anharmonicity extracted from two-tone spectroscopy experiments, plotted as a function of the qubit frequency $f_{01}$. The black data points were extracted from \textbf{a}, while the red stem from larger gate voltages (around $V_{\rm g}\approx -20V$). The top axis labelled $T_{\Sigma}$ is obtained directly from the qubit frequency. The solid lines depict anharmonicity inferred from different models (see main text).  } 
  \label{fig:2}
\end{figure*}

Readout and manipulation of the gatemon qubit is accomplished using standard circuit QED techniques. 
When the qubit frequency $f_{\rm q}$ is strongly detuned from the resonator frequency $f_{\rm r}$, the system is in the dispersive regime \cite{blais2004cavity}. The qubit then causes a shift $\Delta f_{\rm r}$ of the resonator frequency from its bare value $f_{\rm r,b}$, where $\Delta f_{\rm r}=f_{\rm r}-f_{\rm r,b} \approx (g/2\pi)^2/(f_{\rm r,b}-f_{\rm q})$ is determined by the qubit-resonator coupling strength $g$. Figure \ref{fig:1}d shows a measurement of the low power transmission amplitude $\lvert S_{21} \rvert$ plotted as a function of the gate voltage $V_{\rm g}$ and the probe frequency $f_{\rm p}$, plotted as the difference from the bare resonator frequency $f_{\rm r,b}$. The minimum in the transmission signal occurs whenever the probe frequency $f_{\rm p}$ is resonant with $f_{\rm r}$.

The bare resonator frequency $f_{\rm r,b}=7.3527\,$GHz can be directly found at gate voltages $V_{\rm g}>-15\,$V, where the semiconducting part of the NW is close to depletion, so that the Josephson current is negligible and the resonator frequency exhibits no shift. The corresponding internal quality factor of $Q_{\rm i} \approx 2.41\times 10^5$ is found by simultaneously fitting the resonator transmission amplitude and phase in the vicinity of $f_{\rm r,b}$ \cite{probst2015efficient}. At lower gate voltages, the hole density in the Ge segment is increased, resulting in a larger Josephson current on the order of a few tens of nA through the JJ, with a correspondingly larger Josephson energy and an increased qubit frequency. The resonator frequency is Lamb-shifted to higher values accordingly, with a maximum of $\Delta f_{\rm r,max}\approx 6\,$MHz found at $V_{\rm g}\approx -33\,$V, pointed out in Fig.~\ref{fig:1}d. The dispersive shift is positive, suggesting a qubit frequency lower than the bare resonator frequency in this gate voltage range. $\Delta f_{\rm r}$ does not increase monotonically with gate voltage, possibly due to resonances in the semiconducting NW section \cite{ridderbos2019hard}. We also find discontinuities in $f_{\rm r}$ at specific gate voltages, for example at $V_{\rm g}\approx30\,$V. These, we attribute to a switching of the qubit frequency $f_{\rm q}$ due to charge traps in the vicinity of the NW. We note that the resonator frequency shift remains highly reproducible, including the switching, within a gate voltage range of a few volts, but becomes less reproducible when subjected to a wider gate sweep, probably due to rearrangements of such localized charges \cite{luthi2018evolution}.  

To study the quantised energy levels of the device, we perform pulsed two-tone spectroscopy. We apply a $500\,$ns drive tone at frequency $f_{\rm d}$ to the feedline, followed by a $1\,\mu$s probe tone at an optimised frequency of $f_{\rm p}\approx f_{\rm r}\,$ to probe the resonator. We note that we obtain similar results when driving via the side gate. The in-phase $I$ and quadrature $Q$ components of the transmitted probe tone are obtained using heterodyne detection techniques. We plot the normalized amplitude $A/A_0$ of the complex signal, which depends on the state of the qubit. In order to resolve fine features, we plot the numerically calculated second derivative of $A/A_0$ with respect to $f_{\rm d}$, as shown in Fig.~\ref{fig:2}a. The raw data can be found in SI Fig.~\ref{figsup:1}. To avoid large charge rearrangement, we focus on the small gate voltage range of $\sim 600\,$mV shown in Fig.~\ref{fig:2}a, in which the qubit frequency can be tuned between $5.4\,$GHz and $6.3\,$GHz, with a typical qubit-resonator coupling factor of $g\approx 47\,$MHz. We attribute the reproducible discontinuity in the qubit frequency at $V_{\rm g}\approx -30.45\,$V to a nearby charge trap. We also find several gate-voltage independent sharp resonances indicated by the yellow arrows in Fig.~\ref{fig:2}a, possibly due to two-level fluctuators more remote from the gate, for example in the SiO$_2$ substrate, or due to spurious modes in the electromagnetic environment~\cite{lisenfeld2019electric}.

The main features in Fig.~\ref{fig:2}a are two resonances, moving almost in parallel with changing $V_{\rm g}$, clearly visible in the cross section shown in the inset for $V_{\rm g}=-30.7\,$V. Here, the spacing between the two peaks is $\sim 50\,$MHz. Based on the generic energy diagram for a transmon qubit shown in Fig.~\ref{fig:2}b, we attribute the higher frequency resonance to the qubit transition at frequency $f_{01}$ between state $\lvert 0 \rangle$ and $\lvert 1 \rangle$. As we will show, the lower resonance originates from two-photon processes that drive the transition from state $\lvert 0 \rangle$ to the second excited state $\lvert 2 \rangle$. 

To explicitly identify these two peaks, we perform two-tone spectroscopy as a function of frequency and power of the drive tone, as plotted in Fig.~\ref{fig:2}c. At low powers, only a single resonance is found. For large powers, an additional resonance appears at a slightly lower frequency, consistent with a lower probability for higher order two-photon absorption processes. This transition to the second excited state is also found in the Rabi measurement shown in Fig.~\ref{fig:2}d, where we plot the resonator response as a function of the drive pulse duration $\tau_{\rm Rabi}$ and drive frequency $f_{\rm d}$. The drive pulse induces periodic Rabi oscillations between the states $\lvert 0 \rangle $ and $\lvert 1 \rangle$, resulting in the characteristic Rabi chevron pattern, with the qubit frequency $f_{01}\approx 6.325\,$GHz. Again, we find an additional broader feature for longer pulse times at $\sim 6.275\,$GHz, typically attributed to the $f_{02}/2$ resonance \cite{wang2019coherent,antony2021miniaturizing}. We point out that the Rabi experiments and the power dependence were not taken at the same gate voltage, because of a slight drift over several weeks of measurements. 

This level structure now allows us to directly assess the anharmonicity of the gatemon spectrum $\alpha= 2h(f_{02}/2 − f_{01})$, for different gate voltages. The corresponding data are plotted as black and red points as a function of the gate voltage dependent $f_{01}$ in Fig.~\ref{fig:2}e. We note that for some specific frequencies, spurious resonances hindered us from extracting the peak positions. Details about the data extraction are discussed in the corresponding SI section.

Figure~\ref{fig:2}e is one of our main results, which we now use to estimate the transmission probability and the number of conducting channels in the NW junction. Since the semiconducting NW segment is very short ($33\,$nm) compared to typical superconducting coherence lengths, the JJ is in the short-junction limit. The Josephson potential is then well described by $U(\hat{\phi})=-\Delta \sum_{i}\sqrt{1-T_{i} \sin^{2}(\hat{\phi}/2)}$, where $\Delta$, $T_{i}$, and $\hat{\phi}$ are the superconducting gap, the individual channel transparencies, and the phase difference between the left and right Al segments \cite{beenakker1991universal}. The gatemon Hamiltonian then reads $\hat{H}=4E_{c}\hat{n}^2+U(\hat{\phi})$, which can be expanded to fourth order in $\hat{\phi}$ around the potential minimum at $\hat{\phi}=0$, with the non-harmonic terms used as a perturbation to the harmonic oscillator solutions \cite{kringhoj2018anharmonicity}. This procedure yields

\begin{equation}
    hf_{01}=\sqrt{8E_{\rm c}E_{\rm J}}-E_{\rm c}\left(1-\frac{3 \sum_i T_i^2}{4 \sum_i T_i}\right)
    \label{eq:1}
\end{equation}
and
\begin{equation}
    \alpha=-E_{\rm c}\left( 1-\frac{3 \sum_i T_i^2}{4 \sum_i T_i} \right),
    \label{eq:2}
\end{equation}

with the Josephson energy given by the prefactor of the harmonic $\hat{\phi}^2$-term in the Hamiltonian, $E_{\rm J}=\frac{\Delta}{4} \sum_i T_{i}$. Here, we use the gap $\Delta = 210\, \mu$eV found in DC transport experiments in a control device (see SI Fig.~\ref{figsup:DC}). For this device, we estimate $E_{\rm J}/E_{\rm c} \gtrsim 80$, so that we can approximate $hf_{01}\approx \sqrt{8E_{\rm c}E_{\rm J}}$. Using the numerically simulated $E_{\rm c}$ and the measured qubit frequency $f_{01}$, we obtain the total transmission $T_{\Sigma}:=\sum_i T_{i}=(hf_{01})^2 / ( 2E_{\rm c}\Delta )$, used as the top axis of Fig.~\ref{fig:2}e. The measured quantities $hf_{01}$ and $\alpha$ contain $T_{\Sigma}$ and $\sum_i T^2_{i}$, allowing us to estimate the number of channels and the corresponding transmission probabilities. To do this, we consider two limits \cite{kringhoj2018anharmonicity}. First, we assume $N$ channels of equal transparency $\bar{T}=T_{\Sigma}/N$, yielding  the anharmonicity $\alpha=-E_{\rm c}(1-\frac{3}{4}T_{\Sigma}/N)$. Figure \ref{fig:2}e shows the cases $N=2$, $N=3$ and $N\rightarrow \infty$. This equal-transmission model provides an estimate of the number of active channels, with the case of $N\rightarrow \infty$ giving the SIS tunnel junction limit $\alpha=-E_{\rm c}$, with many low-transparency channels. All data points lie below the $N=2$ case, suggesting that at most two channels dominate the JJ transport. The corresponding mean channel transparency is electrically tunable from $\bar{T}\approx 0.55$ to $\bar{T}\approx 0.77$ in this qubit frequency range.

In the second limit, often called the ``quantum point contact (QPC) limit", we assume one fully transmitting ($T_{{\RN{1}}}=1$) and one partially transmitting channel, with a free parameter $T_\RN{2}$. This limit yields the lowest possible $\lvert \alpha \rvert$ for a given $T_{\Sigma}$. The resulting dependence is shown in Fig.~\ref{fig:2}e as the blue solid curve. Since this curve captures our data better than the equal-transparency model, we conclude that the annealed Al-Ge-Al JJ carries one highly transparent channel, consistent with the large values up to $T\approx 0.96$ reported in previous DC transport studies on similar JJs \cite{sistani2019highly,ridderbos2018josephson}, and a second channel with a lower transmission, electrically tuned in the range from $T_\RN{2}\approx 0.1$ to $T_\RN{2}\approx 0.54$.

\subsection{Coherent control}

\begin{figure}[ht]
    \centering
    \includegraphics[width=\linewidth]{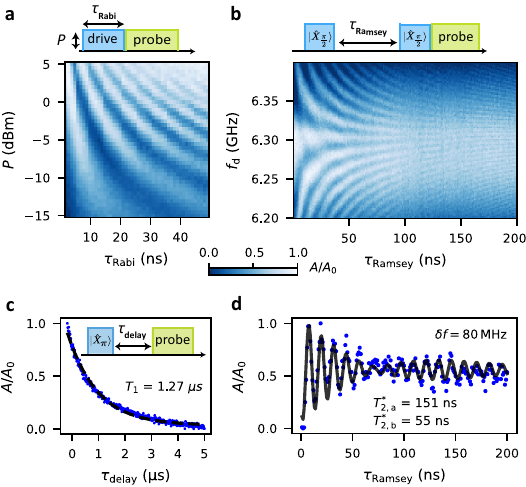}
    \caption{ \textbf{Coherent manipulation of the gatemon.} \textbf{a}, Rabi oscillations as a function of drive power $P$ and pulse duration $\tau_{\rm Rabi}$, measured at $f_{\rm d}=6.345\,$GHz and $V_{\rm g}=-30.7\,$V. \textbf{b}, Ramsey fringes as a function of drive frequency $f_{\rm d}$ and time delay $t_{\rm  Ramsey}$ measured at $V_{\rm g}=-30.6\,$V. Two $X_{\pi/2}$ pulses separated by $\tau_{\rm Ramsey}$ are applied prior to the probe tone. \textbf{c}, Measurement of the characteristic energy relaxation time $T_1$ at $V_{\rm g}=-30.7\,$V. A $10\,$ns $X_{\pi}$ pulse at $P=-10\,$dBm drives the qubit into state $\lvert 1 \rangle$. After a delay time $\tau_{\rm delay}$, the qubit state is readout. The black dashed line shows a fit of an exponential decay, yielding  $T_1=1.27 \,\mu$s. \textbf{d}, Ramsey experiment measured at $V_{\rm g}=-30.6\,$V at a detuning of 80 MHz. Two dephasing times of $T^*_{2,a}$ and $T^*_{2,b}$ are extracted by fitting to a double sinusoidal functions with different exponential envelopes.}
    \label{fig:3}
\end{figure}

To demonstrate the full functionality of the gatemon qubit, we performed time-domain measurements, namely Rabi oscillations and Ramsey interferometry, and extract the corresponding relaxation and dephasing time. For these measurements, we use a Josephson parametric amplifier (JPA) to enhance the readout signal already at base temperature \cite{winkel2020nondegenerate}. In the following experiments, all pulse sequences start with a common initialisation time of $100\,\mu$s to allow the qubit to relax to the ground state $\lvert 0 \rangle$. We plot the normalized resonator response $A/A_{0}$, which essentially represents the occupation of state $\lvert 1 \rangle$. Each data point is averaged over 10000 identical pulse sequences.

First, we perform Rabi experiments at a fixed gate voltage $V_{\rm g}$ and drive at the frequency $f_{\rm d}\approx f_{01}$. The drive pulse of duration $\tau_{\rm Rabi}$ and power $P$ is immediately followed by a $1\,\mu$s probe pulse. The drive pulse induces Rabi oscillations between the states $\lvert 0 \rangle$ and $\lvert 1 \rangle$ defining the z-axis of a Bloch sphere. The resulting qubit-state dependent resonator response is plotted in Fig.~\ref{fig:3}a. As expected, the Rabi oscillations become faster for larger drive powers $P$, which allows for a calibration of the corresponding rotation on the Bloch sphere, around an axis we define as the x-axis. For example, a $X_{\pi/2}$ pulse is obtained as a $5\,$ns drive pulse at $P=-10\,$dBm. We note that we perform these experiments at low powers. If the qubit is driven at larger powers, the two-photon processes become more probable, resulting in a deviation from the linear dependence of the Rabi frequency on the drive amplitude for $hf_{\rm Rabi}\gtrsim \alpha$, and a corresponding leakage from the computational space \cite{krantz2019quantum}, as discussed in SI Fig.~\ref{figsup:rabipopwer}.

Next, we perform Ramsey interferometry, with the corresponding pulse sequence shown schematically in the upper panel of Fig.~\ref{fig:3}b. The qubit state vector is first rotated into the xy-plane of the Bloch sphere using a calibrated $X_{\pi/2}$ pulse with the drive frequency $f_{\rm d}$ sightly detuned from the qubit frequency by $\delta f=f_{\text {d}}-f_{01}$. During a subsequent delay time $\tau_{\text {Ramsey}}$, the qubit state precesses around the z-axis of the Bloch sphere, and a phase of $\phi=2 \pi \delta f \tau_{\text {\rm Ramsey}}$ is accumulated. Then the state vector is again rotated by a $\pi/2$ pulse and the qubit occupation read out as a function of $\phi$. This results in the Ramsey fringes shown in Fig.~\ref{fig:3}b, where the resonator response is plotted as a function of $f_{\rm d}$ and $\tau_{\rm Ramsey}$. The careful timing of these two types of pulses allows one in principle to reach any quantum superposition state on the Bloch sphere.

To quantitatively assess the qubit quality, we now measure the energy relaxation time $T_{1}$ and the dephasing time $T^*_{2}$. First, to measure $T_{1}$, a $X_{\pi}$ pulse is applied to bring the qubit to state $\lvert 1 \rangle$. After a delay time $\tau_{\rm delay}$, the occupation is measured with a probe pulse. Due to the qubit relaxing to the ground state, the probability of finding the qubit in the excited state decays exponentially over a characteristic time scale of $T_1=1.27\,\mu$s. Similarly, we extract the dephasing time $T^*_{2}$ from a Ramsey measurement at a detuning of $\delta f = 80\,$MHz, see Fig.~\ref{fig:3}d. Instead of the expected single frequency with a dephasing time $T^*_2$, we find a beating pattern with two slightly different frequencies, similar to the one found in InAs NW gatemons \cite{luthi2018evolution} and granular aluminium fluxonium qubits\cite{rieger2023granular}.  From a fit to two sinusoidal functions with individual exponential decay times, we obtain the two frequency components $f_{\rm Ramsey,a}=80.17\,$MHz and $f_{\rm Ramsey,b}=86.05\,$MHz. The corresponding time constants are $T^*_{2,{\rm a}}=151\,$ns and $T^*_{2,{\rm b}}=55\,$ns. These two frequencies near $f_{01}$ are also found in the high resolution drive power dependent two tone experiments shown in SI Fig.~\ref{figsup:3}. We tentatively attribute the two frequencies to two qubit configurations determined by a nearby low-frequency two-level fluctuator beyond our experimental control. Further qubit characteristics and a discussion of possible physical mechanisms limiting our qubit performance can be found as supplementary information. A comparison between our qubit performance with other gatemons can be found in Fig.~\ref{figsup:Comparison}, which illustrates that our device exhibits the significantly longer gatemon coherence times than previously reported for group IV materials, comparable to recent experiments on InAs platforms.

\section{Summary and outlook}
In summary, we have demonstrated a fully functional gatemon qubit based on a narrow Ge/Si core/shell nanowire, with a highly transmissive few-channel Josephson junction fabricated using a simple annealing technique, without resorting to sophisticated epitaxial growth techniques. From a detailed analysis of the gatemon anharmonicity, we conclude that the JJ is dominated by two quantum channels with transmissions up to unity. To demonstrate coherent control in the time domain, we performed Rabi and  Ramsey interferometry experiments, yielding a $T_{1}$ time on par with gatemons in the more established III/V materials. Since we find $T^{*}_{2} \ll 2T_{1}$, the qubit coherence is not limited by the energy relaxation, but rather by dephasing caused by on-chip noise sources \cite{lisenfeld2019electric,vepsalainen2020impact}, suggesting that these times can still be improved significantly.

Our experiments show that Ge/Si core/shell NW gatemons are a competitive platform for electrically tunable supercondcuting qubits. JJs in the few-channel QPC limit might significantly reduce the charge noise sensitivity  as shown in Ref.~\cite{reznikov1995temporal,kringhoj2020suppressed,bargerbos2020observation,willsch2023observation}, possibly alleviating the requirement for large-footprint capacitors. Ge/Si core/shell NWs also offer additional design parameters not considered for superconducting qubits so far, namely the exceptionally strong and electrically-tunable spin-orbit interaction of the hole carriers, as well as a tunable Landé $g$-factor \cite{kloeffel2018direct,kloeffel2011strong,froning2021ultrafast}. The fabrication techniques and material system employed here are relevant beyond the gatemon qubits. Other platforms, like Andreev spin qubits \cite{tosi2019spin,hays2020continuous,hays2021coherent,pita2023direct} or Andreev level qubits \cite{zazunov2003andreev,janvier2015coherent,cheung2023photon} will benefit even more from the reduced hyperfine interaction of the hole carriers, and from the sharp, homogeneous interfaces to the Ge islands. Our results opens up new avenues to study Andreev bound states and other, more exotic subgap states in large-spin orbit materials.

\section{acknowledgements}
We thank Tom Jennisken for his assistance in optimizing the annealing recipe, Ioan Pop, Patrick Winkel and Thomas Reisinger from Karlsruhe Institute of Technology for providing the Josephson parametric amplifier, and Hugues Pothier, Marcelo Goffman, Bernard van Heck and Vladimir Manucharyan for fruitful and interesting discussions. This research was supported by the Swiss Nanoscience Institute (SNI) and the Swiss National Science Foundation through grants No 172638 and 192027 and through the NCCR-Spin, grant No 180604. We further acknowledge funding from the European Union’s Horizon 2020 research and innovation programme, specifically a) grant agreement 862046, project TOPSQUAD, b) grant agreement No 828948, FET-open project AndQC, and c) grant agreement No  847471, Marie Skłodowska -Curie, COFUND-QUSTEC.

\section{Methods}
The circuit QED chip was fabricated on an undoped silicon substrate with a $100\,$nm top thermal oxide, using a combination of optical and electron-beam lithography. After cleaning the wafer, a $68\,$nm NbTiN film was sputter deposited on the wafer, in which the resonator, feed line and gate line were patterned using optical lithography, and further defined using dry etching in an ICP-RIE process with Ar and Cl$_2$ gas. Next, we fabricated gold markers to later align the nanowire, and we deposited gold islands (size $\sim \rm 500 \,\mu {\rm m} \cdot 500 \, \mu$m) near the borders of the NbTiN film for better thermalization. The resulting resonator quality factors we obtained in fits shown in SI Fig.~\ref{figsup:s21fit}. 

The Ge/Si core/shell NWs were then transferred to the circuit QED chip using a micro-manipulator. Source and drain contacts as well as side gates were fabricated using standard electron beam lithography. To remove the native oxide on the Si shell, an $8\,$s wet etch in buffered HF (buffered oxide etchant 10:1, 4.6\% HF) was performed, followed by rinsing in DI water. Next, the chip was immediately loaded into the evaporator where $40\,$nm of Al was deposited by thermal evaporation. After lift-off, the chip is annealed on a hotplate in ambient at $200^{\circ}$C for $10\,$min. The NWs were then imaged in a scanning electron microscope to confirm the channel length. Importantly, the annealing and imaging steps can be iterated to obtain the targeted channel length. For each qubit island, we fabricated 3 or 4 annealed junctions, from which we selected the most promising one and connected it to the capacitor and the grounding plane. Ar milling was used to remove the native Al oxide before contacting. The DC and RF wiring of the dilution refrigerator, as well as the measurement setups are discussed in SI Fig.~\ref{figsup:Fridge}. 

\textbf{Author contributions} $\,$H.Z. and L.Y.C. fabricated the device with help from N.S., A.K., J.R., and J.H.U. H.Z., L.Y.C. and N.S. performed the measurements. C.C. helped measurements with parametric amplifier. H.Z., L.Y.C. and N.S. analyzed the data with inputs from A.K., R.H., A.B. and C.S. A.L. and E.P.A.M.B. conducted the nanowire growth. H.Z. wrote the manuscript with helps from all authors. Project was supervised by A.B. and C.S. 

\textbf{Conflicts of interest} $\,$ The authors declare no conflicts of interest.

\textbf{Data availability}  $\,$ All data in this publication are available in numerical form at: \url{https://doi.org/10.5281/zenodo.10198946}

\bibliography{reference}

\pagebreak
\widetext
\begin{center}
\textbf{\large Supplemental to Coherent control of a few-channel hole type gatemon qubit}
\end{center}

\setcounter{equation}{0}
\setcounter{figure}{0}
\setcounter{table}{0}
\setcounter{page}{1}
\makeatletter

\renewcommand{\appendixname}{Supplementary Material}
\renewcommand{\thefigure}{S\arabic{figure}} \setcounter{figure}{0}
\renewcommand{\thetable}{S\arabic{table}} \setcounter{table}{0}
\renewcommand{\theequation}{S\arabic{table}} \setcounter{equation}{0}
\renewcommand{\thesection}{\Alph{section}}

\section{Device overview}  
\begin{figure*}[ht]
    \centering
    \includegraphics[width=\textwidth]{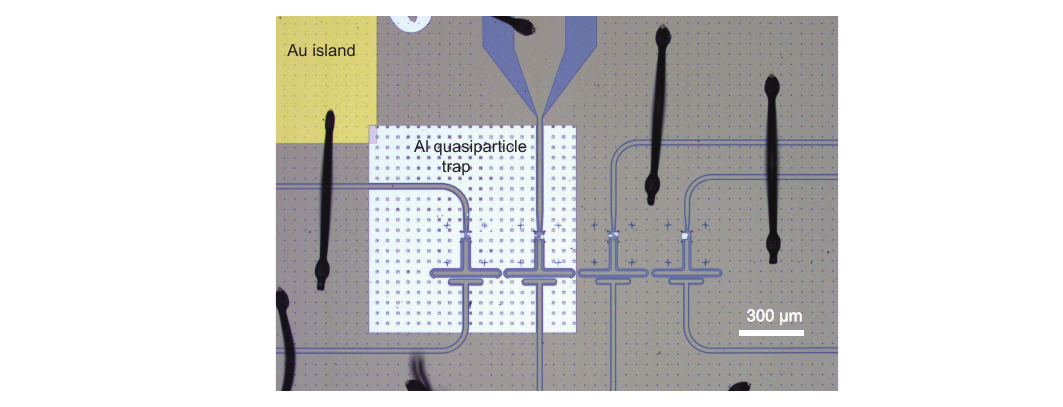}
    \caption{Optical micrograph of the circuit QED chip. Four qubits were fabricated on each chip (center), coupled to the readout resonator (not fully shown). Al was deposited on the NbTiN ground plane as a quasiparticle trap and Au islands for better thermalization of the chip.}
    \label{figsup:Al traps}
\end{figure*}

An optical micrograph of the circuit QED chip is shown in Fig.~\ref{figsup:Al traps}. Four qubits are in the center of the image, capacitively coupled to individual readout resonators (not fully shown). As mentioned in the main text, a $100\,$nm thick Al layer (bright square) was deposited on the NbTiN ground plane (gray background) as quasiparticle trap. Near the borders of the chip, we deposited a several Au islands for better thermalization. 

\section{Peak extraction}  \label{sup:peak}
\begin{figure*}[ht]
    \centering
    \includegraphics[width=\textwidth]{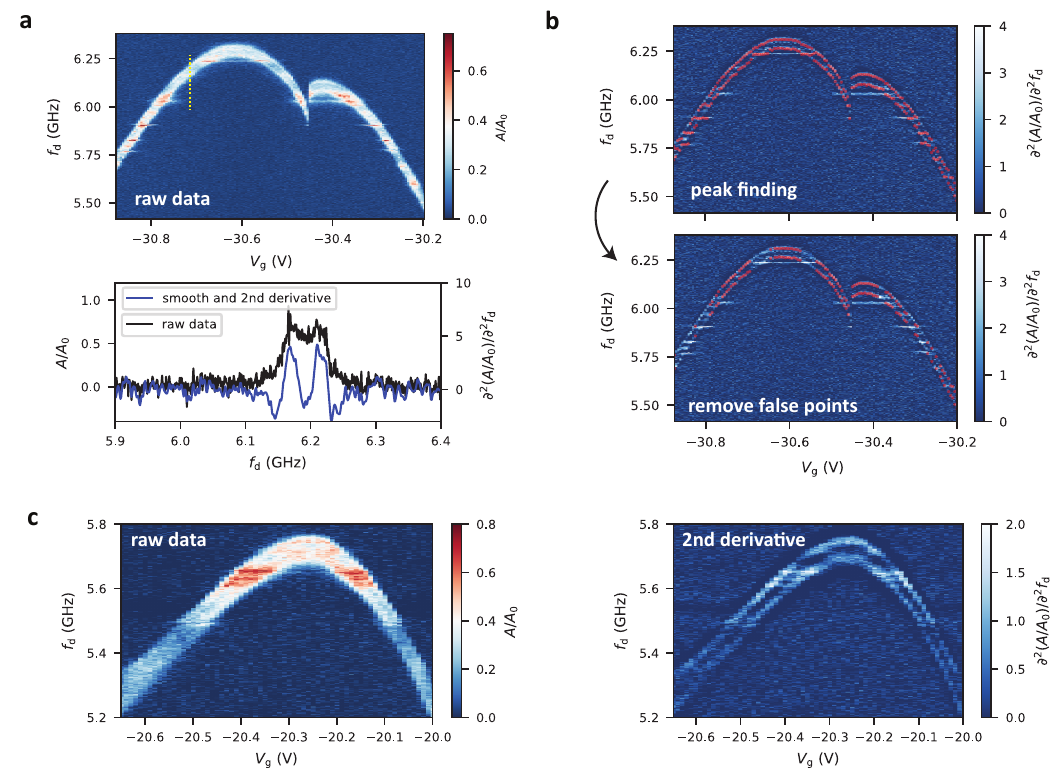}
    \caption{ \textbf{Raw data and peak extraction.} \textbf{a}, Raw data from Fig.~\ref{fig:2}a of the main text, showing the two-tone spectroscopy of the gatemon as a function of the drive frequency $f_{\rm d}$ and the gate voltage $V_{\rm g}$. The indicated cross section (yellow dotted line) of raw data and the corresponding processed data are shown in the bottom panel. \textbf{b}, Extracted peak positions (red dots) superimposed on the two-tone spectroscopy map. Due to gate-independent spurious resonances, a subset of the peaks was disregarded. Only the ones shown in the bottom panel were used. \textbf{c}, Two-tone spectroscopy in a different gate voltage range used for the additional points in the anharmonicity plot of Fig.~\ref{fig:2}e of the main text. Left panel: raw data, right panel: second derivative of smoothed data.}
    \label{figsup:1}
\end{figure*}

In this section, we elucidate how we extracted the peak positions in the two-tone experiments. The raw data of Fig.~\ref{fig:2} (a) is shown in Fig.~\ref{figsup:1}a, with a cross section at $V_{\rm g}=-30.7\,$V in the bottom panel. To better identify the double-peak structure already visible in the raw data, we first applied a moving average with a window size of 10 pixels along the frequency axis, and then took the second derivative. The frequency step size in the raw data is $1\,$MHz. A cross section of the processed data is displayed alongside the raw data in the lower panel of Fig.~\ref{figsup:1}a. 

Next, we extract the peak positions as the positions of the maxima in the second derivative of the smoothed raw data. These positions are pointed out by red dots in the two-tone spectroscopy map in Fig.~\ref{figsup:1}b. We manually discard peaks close to horizontal resonances that we attribute to gate voltage independent two-level fluctuators. From the difference between the two maxima we extract the anharmonicity, as discussed in the main text. Fig.~\ref{figsup:1}c shows additional data and the corresponding second derivative for a different gate voltage range.

\section{Power dependence of the Rabi oscillations}

\begin{figure*}[ht]
    \centering
    \includegraphics[width=0.6\textwidth]{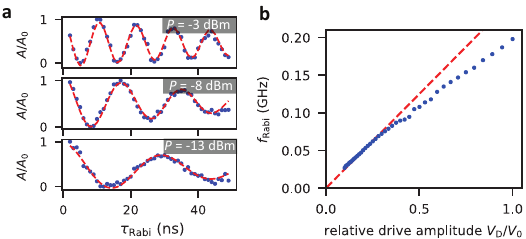}
    \caption{ \textbf{Power-dependent Rabi measurements.} \textbf{a}, Rabi oscillations at different drive powers $P=-3\,$dBm, $P=-8\,$dBm and $P=-13\,$dBm, as indicated. A sinusoidal function with exponential decay (dashed red line) is fitted to the measured data (blue points) to extract the Rabi frequency $f_{\rm Rabi}$ \textbf{b}, $f_{\rm Rabi}$ as a function of the relative drive amplitude $V_{\rm D}/V_{0}$. The red dashed line is a linear fit in the low power range.}
    \label{figsup:rabipopwer}
\end{figure*}
In a pure two-level system, the Rabi oscillations are faster for larger power, with a frequency proportional to the drive amplitude, or to the square root of the power. In Fig.~\ref{figsup:rabipopwer}a, we plot three examples of Rabi oscillations for the indicated (constant) powers. We then extract the Rabi frequency $f_{\rm Rabi}$ by fitting the data points to a sinusoidal function with an exponentially decaying envelope. The resulting $f_{\rm Rabi}$ is plotted as a function of the relative drive amplitude $V_{\rm d}/V_{\rm 0}=\sqrt{P/P_0}$ in Fig.~\ref{figsup:rabipopwer}b. For low amplitudes, $f_{\rm Rabi}$ increases linearly with $V_{\rm d}$, as expected. However, for $f_{\rm Rabi}$ larger than $\sim 80\,$MHz, the observed frequency clearly deviates from the low-amplitude linear dependence. This deviation can be directly attributed to transitions to the second excited state of the gatemon, or in other words, to a leaking out of the computational sub-space. This effect can have two physical origins: 1) a larger power results in a larger occupation of the first excited state and therefore to a larger probability of the two-photon processes that drive the $\lvert 0 \rangle$ $\rightarrow$ $\lvert 2 \rangle$ transition. 2) drive pulses of short duration result in a broader frequency spectrum, if assuming a Gaussian broadened pulse shape \cite{krantz2019quantum},  starting to drive the $\lvert 1 \rangle$ $\rightarrow$ $\lvert 2 \rangle$ transition \cite{krantz2019quantum}. The deviation from the basic low-power linear dependence occurs around $\sim 80\,$MHz, consistent with the anharmonicity found in the main text. This discussion directly illustrates how the anharmonicity limits the qubit operation speed.

\section{Beating pattern in Ramsey experiments}

\begin{figure*}[ht]
    \centering
    \includegraphics[width=0.8\textwidth]{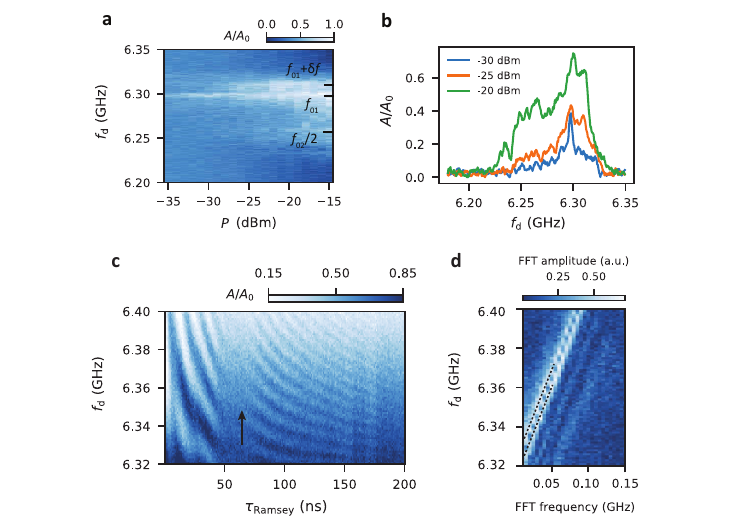}
    \caption{ \textbf{Beating pattern in Ramsey measurement.} \textbf{a}, Two-tone spectroscopy as a function of drive power $P$ and drive frequency $f_{\rm d}$. Specific cross sections at different drive powers are shown in \textbf{b}. \textbf{c}, Example of a Ramsey type measurement described in the main text. The signal vanishes intermittently around $\tau_{\rm Ramsey} \sim 70\,$ns, pointed out by a black arrow, as discussed in the main text. \textbf{d}, Fast Fourier transform of \textbf{c} showing two main qubit frequencies, indicated by black dashed lines.}
    \label{figsup:3}
\end{figure*}

In Fig.~\ref{figsup:3}, a high-resolution power-dependent two-tone spectroscopy experiment is presented. In addition to the resonances at $f_{01}$ and $f_{02}/2$ discussed in the main text, a third transition at a frequency offset $\delta f$ above $f_{01}$ is visible, with $\delta f$ ranging from $5\,$MHz to $10\,$MHz. We tentatively attribute this resonance to a power dependent fast switching between two slightly different values of $f_{01}$, which leads to the beating pattern observed in our Ramsey measurements. Figure~\ref{figsup:3}c shows an example in which the beating is clearly visible. The signal vanishes around a delay time of $\tau_{\rm Ramsey} \sim 70\,$ns, and re-appears for higher values. The Fourier transform of these data is shown in Fig.~\ref{figsup:3}d, in which two different frequency components are found close to the main qubit transition $f_{01}$ (dashed lines).

The switching between two qubit frequencies could have various origins. In our case, two possibilities can be excluded: 1) quasiparticle poisoning \cite{pan2022engineering,aumentado2023quasiparticle} would block one or more transport channels and therefore change the Josephson energy, and with it the qubit frequency. Since there are only two channels dominating the transport in our JJ (see main text), a poisoning would change the qubit frequency by several GHz, not merely by a few ten MHz as found in our experiments. 2) fluctuations in the Cooper pair number \cite{koch2007charge} should occur on the kHz, not on the MHz scale, based on the ratio $E_{\rm J}/E_{\rm c} \approx 80$. After these consideration, We tentatively attribute the switching frequency to a nearby bi-stable charge fluctuator \cite{luthi2018evolution}.

\section{Relaxation and dephasing mechanisms} \label{section:mechanisms}
With changes in the gate voltage (and therefore the qubit frequency), $T_{1}$ changes seemingly randomly between $0.6$ to $1.3\,\mu$s. Similar to \cite{de2015realization,casparis2018superconducting,wang2019coherent}, we cannot find a clear correlation between $T_{1}$ and $f_{01}$ or $d f_{01}/d V_{\rm g}$.  Moreover, the obtained $T_{1}$ values are much lower than $T_{\text{Purcell}}=1/(2\pi \kappa(g/\Delta^2)) \approx 100 \mu$s ($\kappa=1.5\,$MHz and $\Delta=1.5\,$GHZ), suggesting that the qubit relaxation time is not limited by the Purcell decay \cite{houck2008controlling} into the readout resonator. We tentatively attribute the gate dependence of $T_{1}$ to two-level fluctuators weakly coupled to the qubit, resulting in various energy-relaxation channels \cite{klimov2018fluctuations}. This is supported by the finding that after leaving the device in ambient conditions for several weeks, the average $T_{1}$ dropped by a factor of $\sim 2$.  This suggests that with more impurities adsorbed on the nanowire surface, the qubit energy relaxation time $T_1$ drops considerably. 

\begin{figure}[h]
    \centering
    \includegraphics[width=0.65\textwidth]{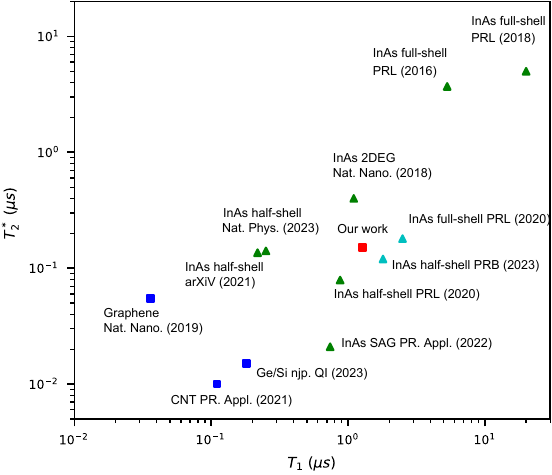}
    \caption{ \textbf{Comaprison of transmon coherence times for different material platforms.} The plot shows literature values for $T_1$ and $T_2^*$ times for various experiments in the literature. Squares symbolize group IV devices, green and cyan triangles stand for III/V InAs-based ones. The cyan data points is a subset of group III/V gatemon experiments in which $T^*_2$ was not reported and we substituted it by $T_{\rm Rabi}$. The data and the corresponding references are listed in Table~\ref{sup:table}}
    \label{figsup:Comparison}
\end{figure}

\begin{table*}[h]
    \centering
    \begin{tabular}{lllllll}
        \hline
        Material & Platform & Superconductor & $T_{1}(\mu s)$ & $T^*_{2}(\mu s)$ & $T_{2,echo}(\mu s)$ & \# channels \\
        \hline
        III/V & InAs NW (full-shell)\cite{casparis2016gatemon} & Epitaxy  & $5.3$ & $3.7$ & $9.5$ & $2$ to $3$ \cite{kringhoj2018anharmonicity} \\
        III/V & InAs NW (full-shell) \cite{luthi2018evolution} & Epitaxy & $20$ & $5$ & $30$ & $2$ \\
        III/V & InAs NW (full-shell) \cite{pita2023direct} & Epitaxy  & $0.251$ & $0.141$ & $0.33$ &  \\
        III/V & InAs NW (full-shell) \cite{sabonis2020destructive} & Epitaxy  & $0.25$ & \textcolor{cyan}{$0.18$} &  &  \\

        III/V & InAs NW (half-shell) \cite{danilenko2023few} & Epitaxy  & $1.8$ & \textcolor{cyan}{$0.12$} &  &  \\
        III/V & InAs NW (half-shell) \cite{uilhoorn2021quasiparticle} & Epitaxy  & $0.218$ & $ $ & $0.136$ &  \\
        III/V & InAs NW (half-shell) \cite{bargerbos2020observation} & Epitaxy  & $0.875$ & $0.079$ & $0.295$ &  \\

        III/V & InAs 2DEG \cite{casparis2018superconducting} & Epitaxy & $1.1$ & $0.4$ & $2.2$ & many \\
        III/V & InAs SAG NW \cite{hertel2022gate} & Epitaxy & $0.74$ & $0.021$ & $1.34$ & $2$ \\
        \hline
        IV & Graphene \cite{wang2019coherent} & Surface & $0.036$ & $0.055$ & & many \\
        IV & Carbon nanotube \cite{mergenthaler2021circuit} & Surface & $0.11$ & ($0.01$) & & $1$ \\
        IV & Ge/Si core/shell thick NW \cite{zhuo2023hole} & Al-Ge exchange & $0.18$ & ($0.015$) & & many \\
        \textcolor[HTML]{CB0000}{IV} & \textcolor[HTML]{CB0000}{Ge/Si core/shell thin NW (our work)} & \textcolor[HTML]{CB0000}{Al-Ge exchange} & \textcolor[HTML]{CB0000}{$1.27$} & \textcolor[HTML]{CB0000}{$0.151$} & & \textcolor[HTML]{CB0000}{$2$} \\
        \hline
    \end{tabular}
    \caption{\textbf{Comparison of gatemon platforms} The columns in this table show the material (group III/V or group IV), the exact platform, the types of superconducting contacts, the best reported qubit characteristics times $T_1$, $T^*_2$, $T_{2,echo}$, as well as the number of channels active in the JJ. Numbers in brackets indicate that $T^*_2$ was not directly measured in the Ramsey experiment, but deduced from the qubit linewidth. For data points in cyan color, $T^*_2$ were not reported in the experiment, we therefore use $T_{\rm Rabi}$ instead. }
    \label{sup:table}
\end{table*}

In our gatemon, $T^{*}_{2} \ll 2T_{1}$, showing that the qubit coherence is not limited by energy relaxation. In addition, we did not find a significant improvement of $T^{*}_2$ on and off the gate sweet spots, for example at $V_{\rm g}\approx -30.6\,$V, nor with better filtering on the DC gate-line by replacing the $80\,$MHz low-pass LC filters by 100 kHz RC filters. From these findings, we conclude that $T^{*}_{2}$ is not limited by gate voltage noise, but rather by on-chip noise originating for example from localized states in the native oxide of the Si shell or in thermal oxide of the substrate, resist residues from fabrication, or from carbon contamination from SEM imaging.

To compare the coherence times of our gatemon to other platforms, we show the results from prior studies in Fig.~\ref{figsup:Comparison} and in Table.~\ref{sup:table}. Our gatemon exhibits the best coherence times of the rather few group IV based platforms to date, on a similar scale as recent InAs based materials with epitaxial Al.

\section{Superconducting gap measurement}
\begin{figure*}[ht]
    \centering
    \includegraphics[width=\textwidth]{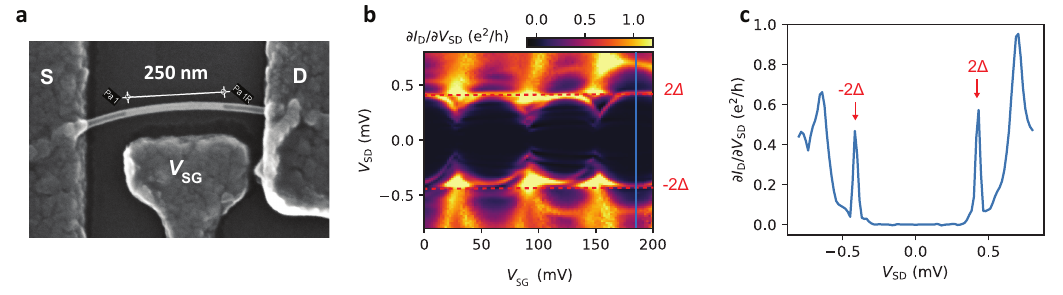}
    \caption{ \textbf{Superconducting gap from DC transport.} \textbf{a}, SEM image of a Ge/Si core/shell nanowire (NW) JJ fabricated with the same annealing method as in the main text. \textbf{b}, DC voltage bias spectroscopy as a function of the gate voltage $V_{\rm SG}$ and the source-drain bias voltage $V_{\rm SD}$. A cross section at $V_{\rm SG}=187\,$mV is shown in \textbf{c}.}
    \label{figsup:DC}
\end{figure*}

The superconducting gap $\Delta$ is measured in a control device, shown in Fig.~\ref{figsup:DC}a. The device is fabricated with the same annealing method, with a Ge channel length of $250\,$nm. The device was cooled down in the same dilution refrigerator as the circuit QED chip.

When the NW is gate-tuned close to pinch off, the NW channel forms a quantum dot, tunnel coupled  to the annealed Al in the NW. In this regime, we can perform tunnel spectroscopy to determine the superconducting gap $\Delta$. For this, we measured the differential conductance $\partial{I_{\rm D}}/\partial{V_{\rm SD}}$ as a function of source-drain voltage bias $V_{\rm SD}$ and side gate voltage $V_{\rm SG}$. As shown in Fig.~\ref{figsup:DC}b, we find Coulomb blockade diamonds with a gap of $2\Delta$ in the transport opened around zero bias. A cross section at the indicated gate voltage is plotted in  Fig.~\ref{figsup:DC}c, where we extract $\Delta \approx  210\,\mu$eV.

\section{Resonator quality factor}

\begin{figure*}[ht]
    \centering
    \includegraphics[width=0.6\textwidth]{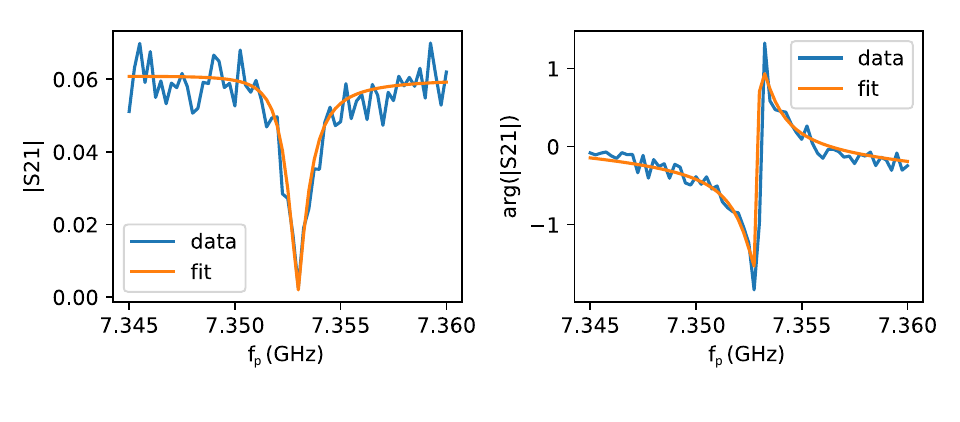}
    \caption{ The measured magnitude and phase of $S_{21}$ in the vicinity of $f_{r,b}$. A simultaneous fit to both quantities using the model in Ref.~\cite{probst2015efficient} results in the internal quality factor $Q_{\rm i} \approx 2.41\times10^5$ and the external quality factor $Q_{\rm c} \approx 4\times10^3$.}
    \label{figsup:s21fit}
\end{figure*}

We use the python based package ``resonator\_tools" \cite{probst2015efficient} to extract the quality factors of our readout resonator. The data and fits of the magnitude and phase of the transmission signal $S_{21}$ are shown in Fig.~\ref{figsup:s21fit}, yielding the internal and external quality factors listed in the caption of the figure.

\section{Measurement setup}
\begin{figure*}[ht]
    \centering
    \includegraphics[width=\textwidth]{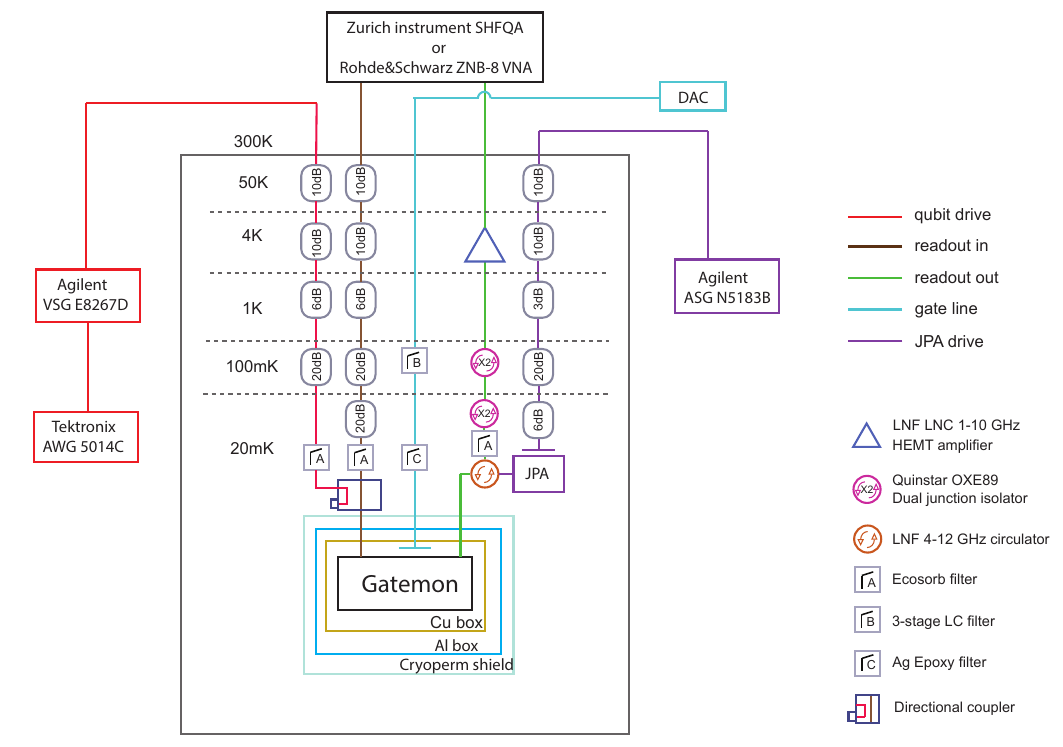}
    \caption{ Schematic of the dilution refrigerator and measurement setup.}
    \label{figsup:Fridge}
\end{figure*}

The RF measurements were performed either using a Rohde\&Schwarz ZNB-8 vector network analyzer or a Zurich Instrument SHFQA Quantum Analyzer. The drive and probe signals were heavily attenuated by 66 dB and filtered with home-made Ecosorb filters at low temperature. The output signals were amplified in an amplification chain consisting of a Josephson parametric amplifier (JPA), several circulators, an Ecosorb filter and dual junction isolators, followed by a HEMT amplifier. All components are shown in detail in Fig.~\ref{figsup:Fridge}.

The qubit drive tone is generated by a vector signal generator (Agilent E8267D) modulated by an envelope signal from an arbitrary waveform generator (AWG, Tektronix 5014C). The DC flux and gate lines are filtered using Ag epoxy filters at the coldplate and a 3-stage LC-filter with the cutoff frequencies $80\,$MHz, $225\,$MHz and $400\,$MHz. The qubit drive via the gate is achieved by combining a DC and an RF line on the PCB with a RC bias tee using a 1 k$\Omega$ resistor and a $15\,$nF capacitor.

\end{document}